\documentclass[10pt,journal,compsoc]{IEEEtran}
\usepackage{amsmath,amsfonts}
\usepackage{algorithmic}
\usepackage{algorithm}
\usepackage{array}
\usepackage[caption=false,font=normalsize,labelfont=sf,textfont=sf]{subfig}
\usepackage{textcomp}
\usepackage{stfloats}
\usepackage{url}
\usepackage{verbatim}
\usepackage{graphicx}
\usepackage{cite}
\usepackage{color}
\usepackage{multirow}

\begin{document}

\title{iSatCR: Graph-Empowered Joint Onboard Computing and Routing for LEO Data Delivery\\

\author{Jiangtao~Luo,~\IEEEmembership{Senior Member,~IEEE}, Bingbing~Xu,~\IEEEmembership{Member,~IEEE}, Shaohua~Xia,~\IEEEmembership{Member,~IEEE}, Yongyi~Ran,~\IEEEmembership{Member,~IEEE}}

\thanks{Jiangtao Luo, Bingbing~Xu, Shaohua Xia, and Yongyi Ran are with the School of Communications and Information Engineering, Chongqing University of Posts and Telecommunications, Chongqing, 400065 China. Jiangtao Luo is the corresponding author. (Email: luojt@cqupt.edu.cn, 1690500735@qq.com, 1805519393@qq.com, ranyy@cqupt.edu.cn). 
}

\thanks{This work is supported by the National Natural Science Foundation of China (No.~U25B2033, No.~62171072 and No.~U23A20275).}
}

\IEEEtitleabstractindextext{\begin{abstract}
Sending massive Earth observation data produced by low Earth orbit (LEO) satellites back to the ground for processing consumes a large amount of on‑orbit bandwidth and exacerbates the space‑to‑ground link bottleneck. Most prior work has concentrated on optimizing the routing of raw data within the constellation, yet cannot cope with the surge in data volume. Recently, advances in onboard computing have made it possible to process data in situ, thus significantly reducing the data volume to be transmitted. In this paper, we present iSatCR, a distributed graph-based approach that jointly optimizes onboard computing and routing to boost transmission efficiency. Within iSatCR, we design a novel graph embedding utilizing shifted feature aggregation and distributed message passing to capture satellite states, and then propose a distributed graph-based deep reinforcement learning algorithm that derives joint computing-routing strategies under constrained on-board storage to handle the complexity and dynamics of LEO networks. Extensive experiments show iSatCR outperforms baselines, particularly under high load.
\end{abstract}

}

\maketitle

\IEEEpubidadjcol

\section{Introduction} 
 \IEEEPARstart{L}{ow} Earth orbit (LEO) satellites will play an increasingly important role in Earth observation owing to their high accuracy, lower payload costs, reduced communication latency, and smaller user terminals. As hundreds or even thousands of LEO satellites have been launched, they will facilitate high-resolution Earth observations\cite{zhang2022progress}. These satellites are finding more and more critical applications in environmental monitoring\cite{lechner2020applications}, agriculture management\cite{klein2021application}, disaster recovery, and urban planning\cite{spireGlobal}. Recently, advances in sensing technologies, particularly hyperspectral imaging (HSI)\cite{ghamisi2017advances}, have significantly improved data precision while dramatically increasing its volume. In such a system, each observation satellite generates approximately 1 TB of data per day~\cite{tao2023transmitting}, which is to be transmitted to the Earth in the format of raw data. Unfortunately, they are not always received on time due to the large amount of bandwidth required and the uneven distribution of ground stations. Consequently, effectively transmitting observation data to the ground has emerged as a critical challenge in Earth observation.

\IEEEpubidadjcol
 
To address the above problem, considerable efforts have been dedicated to exploring more efficient data delivery mechanisms. A mainstream approach is to design optimized routing strategies to improve throughput and reduce latency in LEO satellite networks\cite{lai2021orbitcast, zuo2022deep}. Actually, the routing algorithms for a massive LEO constellation have become much more complicated. Most importantly, such approaches cannot keep up with the rapidly increasing volume of Earth observation data since they do not essentially reduce the volume of data to be transmitted. Alternatively, the development of onboard computing capabilities offers the potential to reduce the transmission burden by processing data onboard before transmission\cite{zhang2022progress}. For instance, data can be compressed onboard to reduce the required transmission bandwidth\cite{wang2022review}\cite{leyva2023satellite}. Furthermore, tasks such as object detection\cite{sowmya2017remote}, disaster forecasting, and environmental parameters extraction\cite{yuan2020deep} require transmitting only a few bytes of results if computed onboard. The advantages of onboard processing effectively address the limitations of routing-only approaches. Thus, in this paper, routing and onboard processing (i.e., computing) are jointly explored to accelerate the utilization of Earth observation data. 

However, three key challenges are faced when jointly optimizing routing and computing. First, high overhead in information collecting and fusion. The topology of a LEO satellite network is subject to constant changes due to factors such as high-speed movement, solar flares, solar storms, Doppler shifts, etc.~\cite{yong2022high}. In addition, the traffic distribution within the network also exhibits significant dynamics due to the time-varying coverage of satellites. Therefore, the network states need to be known in real-time to make decisions in time. However, the frequent collection of global information on computing and network resources incurs significant overhead. Second, selecting the optimal computing sites in time under the uneven and dynamic distribution of computing resources. Due to limited and unevenly distributed onboard computing resources, observation data may not be processed immediately by the local satellite and must be offloaded to other satellites instead. Therefore, routing decisions must take into account whether the satellites along the routing path can meet the onboard processing requirements for the offloaded data. Third, high complexity in joint optimization of routing and computing. On the one hand, large-scale LEO satellite networks, consisting of hundreds to thousands of satellites, present high-dimensional state spaces. On the other hand, the joint decision-making of computing satellites and transmission paths leads to a high-dimensional action space. These two factors can lead to high computational complexity, potentially resulting in the ``curse of dimensionality''. 

To address these challenges, we propose an intelligent satellite computing and routing scheme, termed \textit{iSatCR}, a deep
reinforcement learning algorithm that improves the transmission efficiency of Earth observation data by jointly optimizing computing and routing in LEO satellite constellations. In iSatCR, a novel graph embedding mechanism with shifted feature aggregation and distributed message passing is designed to achieve efficient sensing of satellite states. Given the high complexity and dynamics of the LEO satellite network, we propose a distributed, graph-empowered deep reinforcement learning method to optimize computing and routing strategies for Earth observation tasks under limited satellite storage. Finally, extensive experiments demonstrate that iSatCR outperforms baseline algorithms. The main contributions of this work are outlined as follows:

\begin{enumerate}[]
            
    \item Design a novel graph embedding-based sensing mechanism for each satellite to gather and share computing and network information from 3-hop neighbors in a distributed manner. The graph embedding consists of a new feature aggregation method called \textit{shifted feature aggregation} and a distributed message passing mechanism, which can precisely preserve spatial information of resources by aggregating the features of satellites in different ranges separately. This sensing mechanism alleviates overhead and effectively achieves resource awareness. 
    
    \item Propose a distributed graph-empowered \textit{dueling double deep Q-Network} (D3QN) method to obtain the joint decision of computing and routing. The graph-empowered approach enhances generalization and adaptability to changes in network topology. The joint optimization problem is modeled as a \textit{partially observable Markov decision process} (POMDP) to minimize task delays under storage constraints. D3QN is used to solve the problem, addressing high computational complexity and capturing the dynamics of the LEO satellite network. Additionally, a heuristic exploration strategy is used to accelerate algorithm convergence.
    
    \item Develop a system-level simulation environment to model the observation missions at the constellation scale. Different algorithms are evaluated under the same random traffic and network conditions. Extensive simulation results demonstrate that iSatCR outperforms baseline algorithms in terms of average delay and packet loss rate, especially under high-load conditions.
    
\end{enumerate}
        
The rest of this paper is organized as follows. First in Section \ref{sec:related}, the related studies are summarized. Then, Section \ref{sec:model} presents the network, task, and delay models used in this work. Section \ref{sec:overview} outlines the proposed iSatCR method. Next, Section \ref{sec:method} introduces the resource awareness mechanism based on a graph embedding method. Then, Section \ref{sec:drl} proposed a distributed decision-making mechanism for joint computing and routing based on D3QN. Next, Section \ref{sec:complexity} analyzes the complexity of the proposed algorithm. Simulation results and analyses are given in Section \ref{sec:result}. Finally, Section \ref{sec:conclusion} concludes this paper. 

\section{Related work}
\label{sec:related}

\subsection{Routing Strategies for LEO Satellite Networks}

Routing strategies in satellite networks are generally categorized into two types: centralized and distributed approaches. Centralized approaches often use time-varying graphs to model the global network state and select routing paths within the entire graph. In contrast, distributed approaches primarily rely on information from local and neighboring satellites to determine the routing path.

Centralized approaches often require the use of complex, time-varying graph models and information-sensing techniques to characterize dynamic topologies of LEO satellite networks. Tang et al. in\cite{tang2018multipath} proposed source-based and destination-based multipath cooperative routing algorithms, which dynamically deliver different parts of a data flow along multiple link-disjoint paths. \cite{zhang2019stag} and \cite{zhang2020application} proposed routing schemes that ensure service transmission QoS using a \textit{time aggregated graph} (TAG). \cite{li2024dynamic} proposes Dyna-STN, a dynamic discrete topology-oriented wide-area routing mechanism that characterizes the time-varying topology of satellite networks using a dynamic discrete topology model. Establishing complete dynamic topologies introduces significant overhead, making centralized approaches less suitable for highly dynamic and failure-prone satellite networks.

Distributed methods mainly rely on information from a few neighboring satellites for routing decisions. \cite{zuo2022deep} and \cite{Ran2025Fully} introduced distributed routing algorithms using deep reinforcement learning (DRL), where each satellite serves as an agent that observes information from its single-hop neighbors. Other distributed approaches~\cite{guo2015weighted,qi2020distributed,zhang2021aser} model the unique topology of satellite networks as grid graphs. These studies used satellite indices within the grid graph to enhance distributed resource awareness and incorporated fault-tolerant mechanisms to adapt to dynamic connectivity changes. However, these methods have limited ability to expand information sensing and adapt to dynamic topology changes.

\subsection{Joint Optimization of Computing and Transmission} 

In the field of satellite onboard computing, most research has primarily focused on leveraging satellites as edge computing nodes to facilitate offloading for ground missions. Zhang et al.\cite{zhang2019satellite}, Ding et al.\cite{ding2021joint}, and Tang et al.\cite{tang2021computation} developed satellite mobile edge computing (MEC), where satellites provide MEC services via satellite links in the absence of nearby ground servers for user devices. Waqar et al.~\cite{waqar2022computation} presented an air-ground connected MEC-supported vehicular network and introduced a distributed value iteration-based reinforcement learning approach to minimize overall computation and communication overhead. Work in \cite{chen2024qos} proposes a distributed QoS-aware offloading algorithm for Internet of things (IoT) devices in LEO satellite edge computing, formulated as a non-cooperative game to minimize total QoS costs under multiple constraints. Gao et al.\cite{gao2024satellites} proposes a computation offloading algorithm based on hierarchical dynamic resource allocation, combining breadth-first search and greedy strategies to optimize computation offloading and resource allocation in satellite edge computing, minimizing delay and energy consumption. Another work~\cite{Chen2025game} proposed a game theory-based distributed computing offloading algorithm. This algorithm can effectively reduce the offloading cost while meeting the resource requirements and the time constraints of low-earth orbit satellite communication. Work in \cite{jiang2025satellite} focuses on the joint optimization of communication, computing, and caching resources, proposing a multi-agent federated reinforcement learning method to minimize the total delay for mobile users in satellite edge computing. Although these studies optimized the allocation of various resources, they primarily focused on terrestrial tasks and only considered satellites providing offloading services for their coverage areas, neglecting the allocation challenges of onboard computing tasks at the scale of the entire constellation.

Work in \cite{huang2024integrated} identifies most of the existing studies related to satellite edge cooperation have focused on small-scale task scheduling and resource allocation among adjacent 
nodes. To address these challenges, it proposes an integrated computing and networking framework for LEO satellite mega-constellations. 
Gong et al.\cite{gong2024intelligent} proposed an edge-intelligence-driven collaborative architecture, integrating sensing, communication, computation, caching, and intelligence services in satellite networks, supported by a centralized and distributed learning framework. Cao et al.\cite{cao2023computing} proposed a computing-aware routing method aimed at integrating onboard computing and routing optimization. Work in \cite{wang2022cdmr} proposed a Computing-Dependent Multi-path Routing (CDMR) paradigm, optimizing energy consumption while ensuring processing latency constraints during task transmission. Guo et al. \cite{guo2025enabling} proposed a joint computing and transmission strategy for real-time satellite network applications, using a k-shortest path algorithm and a graph-based approach to reduce algorithm complexity. However, these centralized approaches may introduce additional latency in decision-making and increase overhead in information sensing, potentially limiting performance in the highly dynamic scenarios of satellite networks.

\subsection{Graph Based DRL Methods in Communication}
    
 In the context of resource allocation within a satellite constellation, the system can be represented as a graph, with satellites as nodes and inter-satellite links (ISLs) as edges. Exploring this inherently non-Euclidean graph structure poses a significant challenge due to its complex topologies and interdependent relationships. In recent years, graph-based reinforcement learning techniques have been used to navigate the complexities of graph structures and address challenges in communication network problems\cite{jiang2022graph}. These include graph neural network methods such as \textit{graph convolutional network} (GCN)\cite{kipf2016semi}, \textit{message passing neural network} (MPNN)\cite{gilmer2017neural}, and \textit{graph attention network} (GAT)\cite{velivckovic2017graph}. These methods have demonstrated notable success in areas such as power control\cite{Wang2024ENGNN}, traffic forecasting\cite{Patil2025Travel}\cite{Peng2025Spatiotemporal}, task offloading\cite{xu2024transedge}, and routing\cite{Zhou2025Distributed}\cite{He2025Routing}. However, as the graph size grows, implementing graph-based methods at the entire graph scale introduces significant computational challenges. As a result, these methods are primarily used in smaller-scale terrestrial networks with less dynamic topologies. These methods are less effective in highly dynamic, large-scale satellite networks.

In contrast, graph embedding methods, such as GraphSAGE~\cite{hamilton2017inductive}, provide a more efficient solution by generating embeddings based on local subgraph features. For instance, Sun et al. in\cite{sun2021mobile} employed GraphSAGE to analyze edge features in IoT network intrusion detection, while work in\cite{Ji2025Graph} utilized GraphSAGE to integrate the local features of nodes with the information of their neighbors, assisting intelligent agents in evaluating the quality of different sub-channels. However, existing graph embedding techniques, such as deepwalk\cite{perozzi2014deepwalk} and structure2vec\cite{dai2016discriminative}, face limitations in aggregating node-specific features, while GraphSAGE struggles to represent precise node-level features and positional information. As a result, these methods are often limited to detection and classification tasks and are unsuitable for resource allocation in LEO satellite constellations.

\section{System Model and Problem Formulation}
\label{sec:model}

\begin{figure}[!t]
    \centering
    \includegraphics[width=0.475\textwidth]{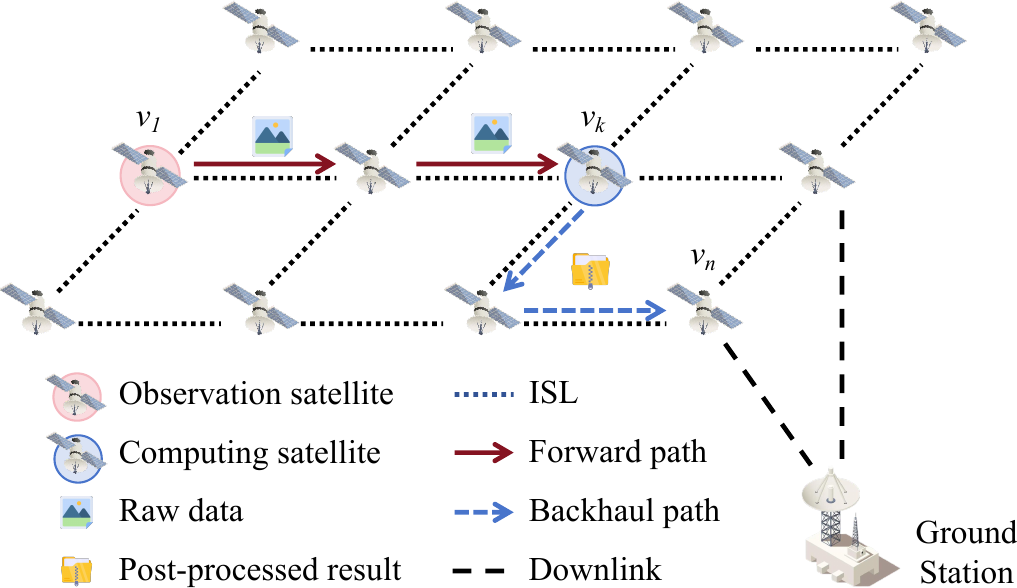}
    \caption{Scenario of an observation task to be computed onboard and results transmitted back to the ground.}
    \label{fig1}
\end{figure}

As shown in  Fig.~\ref{fig1}, the system reference consists of a LEO satellite constellation and ground stations. In this system, after the observation satellites collect data, they transmit the data to the computing satellites for calculation. Then, the processed data is transmitted to the satellite directly connected to the ground station, and finally, it is returned to the ground station by the satellite. Our goal is to design an efficient distributed joint computing and routing strategy for the low Earth orbit satellite constellation to accelerate the utilization of Earth observation data. However, how to select the optimal computing node in real time under the condition of uneven and dynamic distribution of computing resources has become an urgent problem to be solved. Based on the above considerations, we will introduce the network model, task model and delay model next, and construct a joint optimization problem for computing and routing, whose objective is to minimize the system delay including transmission, propagation, queuing and computing delays.

\subsection{Network Model}

A constellation of LEO satellites forms a network with a bidirectional graph structure $G=\{V, E\}$, where $V$ denotes the set of satellites and $E$ the active inter-satellite links (ISLs). $V = \{v_{1}, v_{2}, ..., v_{N}\}$ represent the satellites, with each satellite $v_i$ characterized by its state at any time $t$ as computing capacity $c_i$ in floating point operations per second (FLOPs), occupied storage $m_i(t)$, and queue length for computing tasks $q^c_i(t)$.

The edges $E$ comprise directed ISLs between satellites. An edge $e_{i,j} \in E$ represents a directed link from $v_{i}$ to $v_{j}$, characterized at time $t$ by the link rate $r_{i,j}$, transmission queue length $q^t_{i,j}$, and delay $D_{i,j}(t)$.

\subsection{Task Model}

Due to limited onboard computing resources and hardware memory constraints, along with the input size limitations of deep learning models, satellite observation tasks must be decomposed into multiple parallel subtasks. These subtasks are allocated to different computing satellites for concurrent processing. The raw observation data is divided into overlapping blocks, with each block processed as an independent subtask. These subtasks, such as image preprocessing or on-orbit object detection, are treated as separate data streams. The results are transmitted via inter-satellite links to the designated ground satellite. Once all subtasks are received by the ground server, the data is stitched or fused based on the position annotations of each task’s output.

An observation subtask for processing and transmission in LEO satellites can be described by a tuple $(s, d, s')$, where $s$ is the initial data size, $d$ is the computing demand, and $s'$ is the post-processing data size.

These tasks can be divided into two primary categories: \textit{compression} and \textit{inference}. Compression tasks focus on reducing the raw data size and generally require lower computing power. The compressed data $s'$ is significantly reduced in volume and will be transmitted to ground stations. Inference tasks involve applying machine learning models for target recognition or disaster detection functions, demanding higher computing resources. The results of inference $s'$ are minimal, often at most a few kilobytes.

\subsection{Delay Model}

As shown in Fig.~\ref{fig1}, assuming a observation task is generated at $v_{1}$, with the destination satellite being $v_{n}$, the pre-computation transmission path $P_{1}=\{v_{1},...,v_{k}\}$, the computing satellite being $v_{k}$, the backhaul path being $P_{2}=\{v_{k},...,v_{n}\}$, the task arrival time at each satellites are $t_{1}$, $t_{2}$, ..., $t_{k-1}$, $t_{k+1}$, ..., $t_{n}$ excluding the computing satellite $v_{k}$, $t_{k}$ represent the time when the task finishes computation at satellite $v_{k}$, and $t'_{k}$ is the time when the task arrive $v_{k}$. Assuming the speed of light is $\nu$. Thus, the task process involves four types of delays in the system:

\textbf{Propagation delay} $T_{p}$, which is primarily determined by the distance of each link:

\begin{equation}
	T_{p}=\sum_{i=1}^{n-1} \frac{D_{i,i+1}(t_{i})}{\nu} + \frac{D^{g}_{n}(t_{n})}{\nu}
\end{equation}
where $D^{g}_{n}$ represents the distance from $v_n$ to the connected ground station.

\textbf{Transmission delay} $T_{t}$, which is determined by the task data volume and link rate:

\begin{equation}
	T_{t}=\sum_{i=1}^{k-1} \frac{s}{r_{i,i+1}}+\sum_{i=k}^{n-1} \frac{s'}{r_{i,i+1}}
\end{equation}

\textbf{Queuing delay} $T_{q}$, which includes queuing delays for both computing and transmission, determined by the transmission queue of each link and the computing queue:

\begin{equation}
	T_{q}=\sum_{i=1}^{n-1} \frac{q^{t}_{i,i+1}(t_{i})}{r_{i,i+1}}+ \frac{q^{c}_{k}(t'_{k})}{c_{k}} + \frac{q^{g}_{n}(t_{n})}{r^{g}_{n}}
\end{equation}

where $r^{g}_{n}$ represents the downlink rate from $v_n$ to the ground station and $q^{g}_{n}(t_{n})$ represents the queue length of downlink in $v_n$ at time $t_{n}$.

\textbf{Computing delay} $T_{c}$, which is determined by the computation requirements of the task and the satellite computing capacity:

\begin{equation}
	T_{c}=\frac{d}{c_{k}}
\end{equation}

\subsection{Problem Formulation}
In the transmission of observation tasks, we aim to minimize the total task delay by allocating transmission paths $P_{1}, P_{2}$ and computing satellites $v_{k}$. Furthermore, considering the constraints on on-board storage capacity, we formulate the optimization problem as follows:

\begin{equation}
	\begin{aligned}
		\min_{\substack{P_{1}=\{v_{1},...,v_{k}\}, \\ P_{2}=\{v_{k},...,v_{n}\}}} \quad \sum_{i=1}^{n-1} \frac{D_{i,i+1}(t_{i})}{\nu} + \frac{D^{g}_{n}(t_{n})}{\nu} + \sum_{i=1}^{k-1} \frac{s}{r_{i,i+1}} + \frac{d}{c_{k}} \\
		+ \sum_{i=k}^{n-1} \frac{s'}{r_{i,i+1}} + \sum_{i=1}^{n-1} \frac{q^{t}_{i,i+1}(t_{i})}{r_{i,i+1}} + \frac{q^{c}_{k}(t'_{k})}{c_{k}} + \frac{q^{g}_{n}(t_{n})}{r^{g}_{n}}
	\\
	\begin{aligned} \text{s.t.} \quad & v_{1}, v_{2}, \ldots, v_{k} \in \{v_{i} \mid m_{i}(t_{i}) + s \leq M \}, \\
	 & v_{k+1}, \ldots, v_{n} \in \{v_{i} \mid m_{i}(t_{i}) + s' \leq M\}, \\
	& v_{n} \in \{v_{i} \mid D^{g}_{i}(t_{i}) \leq D_{max}\}
	\end{aligned}
	\end{aligned}
\end{equation}

The constraints ensure that using storage $m_{i}(t)$ on satellites $v_{i}$ along the transmission paths must not exceed the maximum storage resource value $M$. Additionally, the destination satellite $v_{n}$ must be within the coverage area of any ground station. $D_{max}$ represents the maximum allowable distance for the satellite-to-ground link, calculated based on the orbital altitude and the beam coverage angle.

\section{Proposed Approach of Integrated Computing and Routing}

\label{sec:overview}
    \begin{figure}[t]
    \centering
    \includegraphics[width=0.48\textwidth]{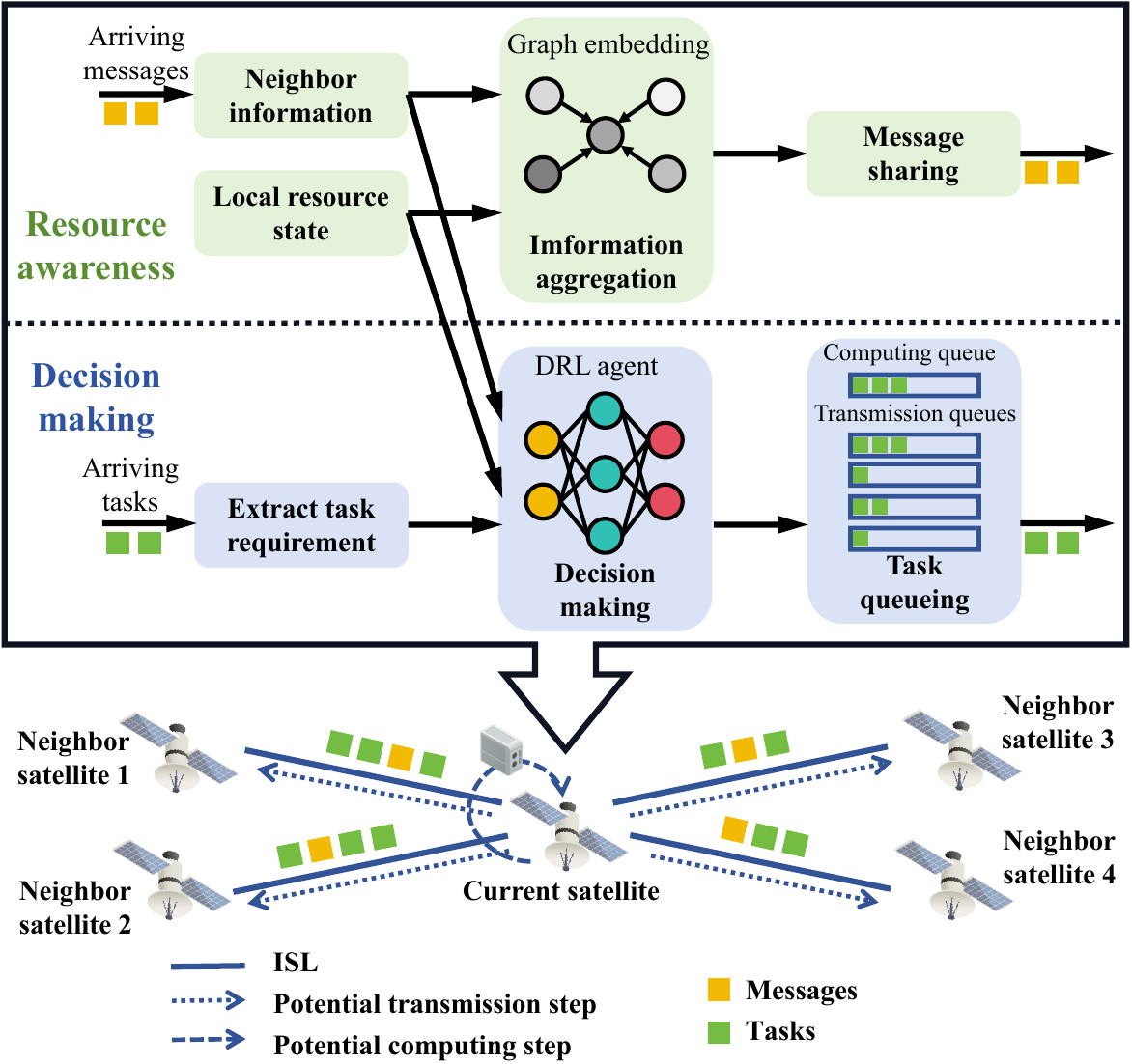}
        \caption{Distributed Framework of joint decision for computing and routing.}
    \label{fig:system_framework}
\end{figure}

This paper introduces a fully distributed intelligence framework designed for the integrated computing and routing of observation missions within an LEO satellite constellation. The framework, depicted in Fig.~\ref{fig:system_framework}, is comprised of two components: resource awareness and decision-making of tasks. In the resource awareness part, satellites receive regular messages from neighboring satellites, aggregate them to generate the resource representation, and share it with neighbors. In the task processing part, the satellite allocates tasks based on resource information along with the task state. Following the output action of the DRL agent, the satellite puts tasks into the corresponding transmission or computing queue. Ultimately, after multi-step transmission, tasks reach the destination satellite and are transmitted to the ground station.

\subsection{Distributed Framework of Joint Decision for Computing and Routing}
The framework consists of two modules: \textit{resource awareness} and \textit{decision making}.

\subsubsection{Resource Awareness}
The resource awareness module shares two main types of information: network topology and resource load data. The network topology refers to the connectivity status within the satellite constellation. The relative motion between satellites causes periodic changes in links, which can be calculated locally at each satellite\cite{pan2019opspf}. Thus, only exceptional link statuses need to be updated and propagated, minimizing communication overhead. In contrast, resource load data refers to the real-time status of computing, transmission, and storage resources on a satellite. Unlike the stable network topology, resource load data undergoes dynamic changes due to network traffic and task strategies, making it more volatile and unpredictable. To address this difference, network topology updates are performed through global flooding, ensuring consistent topology awareness across the network during sudden link state changes, thus preventing destination loss or routing loops. Meanwhile, resource load data is shared periodically between neighboring satellites, ensuring real-time synchronization of load states and enabling dynamic network load adjustment.

\subsubsection{Decision Making}

In the decision making module, the core idea is to treat task processing as a multi-step state transition process, where each satellite only needs to make decisions regarding the next step of the current task. This step-wise decision mechanism effectively reduces computational complexity and enhances the flexibility of task handling. As shown in Fig.~\ref{fig:system_framework}, upon arrival at a satellite, the task enters a waiting queue for the next assignment decision. The task processing strategy is determined by a deep reinforcement learning (DRL) agent, which, based on task requirements and load state information, decides the next destination of the task. Once the decision is made, the task is assigned to the appropriate transmission or computation queue, completing the current state transition. Through collaboration among multiple satellites, the task is transmitted to the target satellite, where necessary computations occur during transmission, and the final results are sent to the ground station. This distributed strategy allows satellites to quickly adjust based on real-time information in response to sudden events, ensuring task processing efficiency and the dynamic adaptability of the network.

\subsection{Graph Embedding For Resource Awareness}
\label{sec:method}

\begin{figure*}[ht]
    \centering
        \subfloat[]{
        \includegraphics[width=0.5\textwidth]{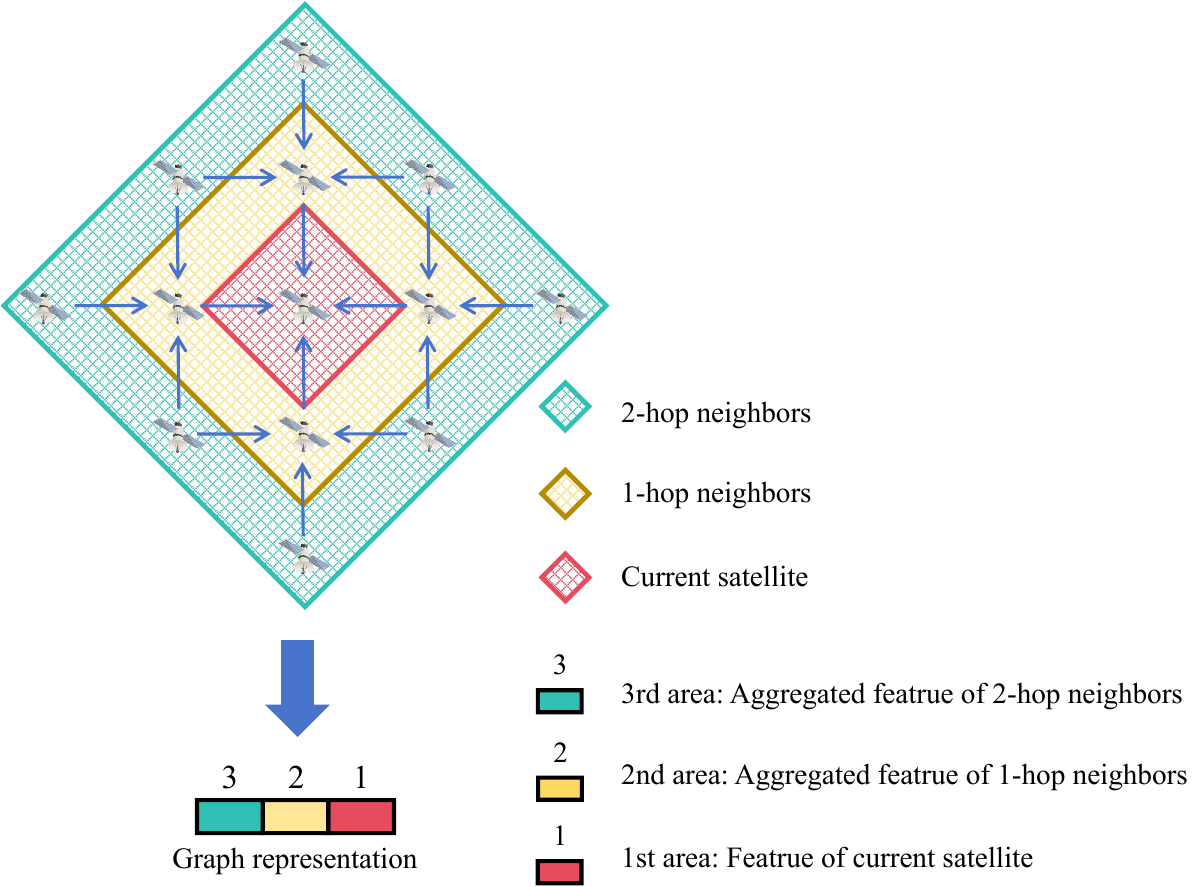}%
        \label{fig:graph2feature}
        }
        \subfloat[]{
        \includegraphics[width=0.4125\textwidth]{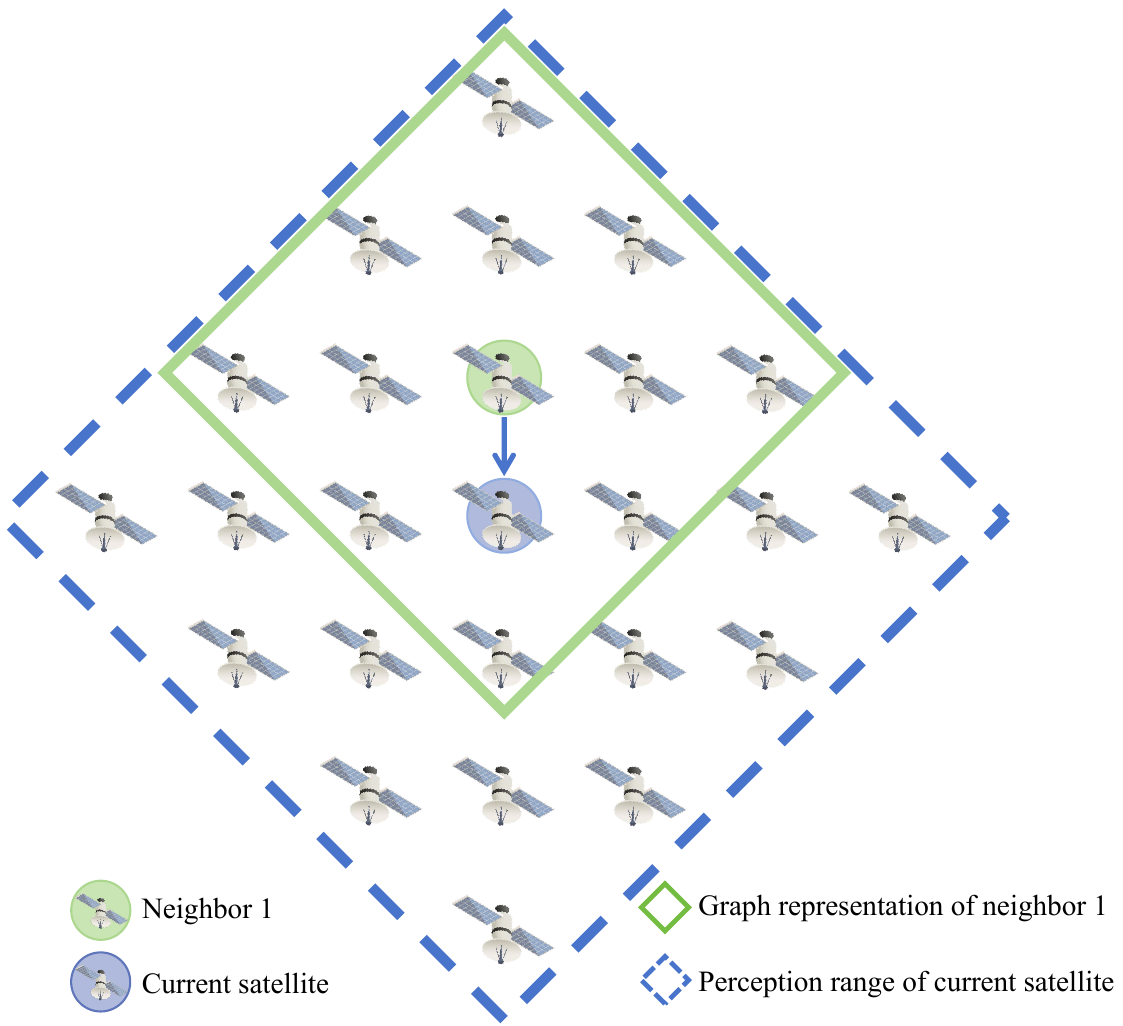}%
        \label{fig:perception_range}
        }
    \hfill
        \subfloat[]{
        \includegraphics[width=0.8\textwidth]{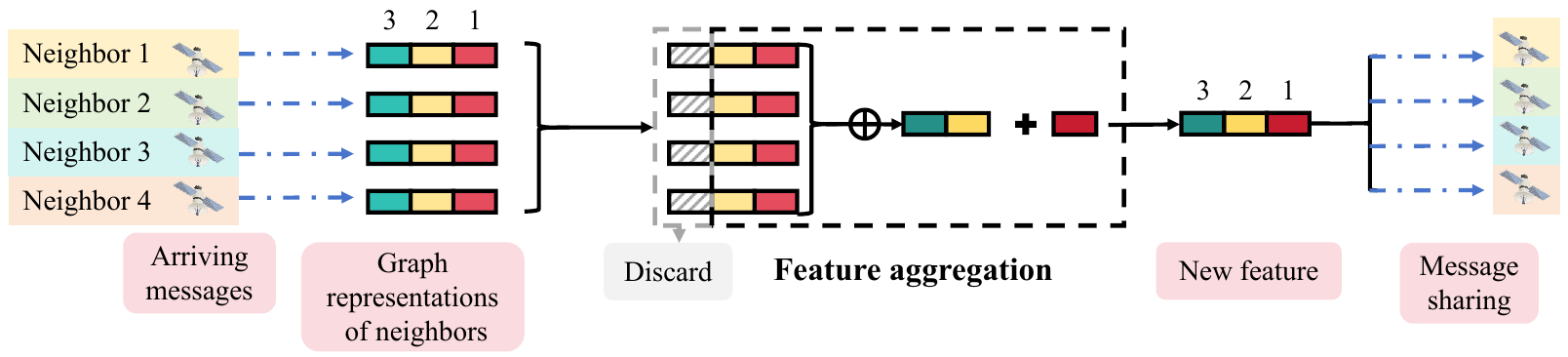}%
        \label{fig:graph_aggrevate}
        }
    \caption{Proposed distributed graph embedding method. (a) Different regions of the graph representation correspond to the current satellite feature, neighboring satellites, and 2-hop neighboring satellites, respectively. (b) An example of the distributed resource awareness mechanism, in which satellites can perceive resource information within a 3-hop range by integrating the graph representations of neighboring satellites. (c) Message passing and feature aggregating of graph representations in distributed graph embedding mechanism.}
    \label{fig:graph_embedding}
\end{figure*}

A graph embedding-based sensing mechanism is proposed to efficient gather and share the dynamic computing and network information of neighboring satellites.  
\subsubsection{Feature Representation of Satellite Resource}
    \label{sec:representation}
	In the LEO constellation, each satellite can be considered as a node in the graph, with its resource information represented by a vector $h^1_{i}$, where $h^1_{i}=\{p_{i},m^{r}_{i},q^{c}_{i},q^{t}_{i}\}$. Here, $p_{i}$ is a binary variable that indicates whether the current node is generating observation data. $m_{r}^{i}$ represents the normalized remaining storage capacity, denoting the proportion of remaining storage resources to the total storage capacity of the satellite. $q^{c}_{i}$ is the current length of the computing queue, which is calculated by dividing the total computation demand of the queueing tasks by the satellite computing capacity. This represents the computation time required for the remaining tasks. $q^{t}_{i}$ represents the normalized total length of transmission queues in all directions. It is calculated based on the proportion of data size in transmission queues to the satellite’s total storage capacity. During feature aggregation, the aim is to understand the current queue situation for computing and transmission resources to minimize task delay. For storage resources, it is important to know the remaining resource capacity to avoid shortages of storage space. Therefore, computing and transmission resources $q^{c}_{i}$ and $q^{t}_{i}$  are represented by current usage values, while storage resources $m^{r}_{i}$ are represented by remaining values, allowing for meaningful feature information after aggregation.
 
\subsubsection{Shifted Feature Aggregation Based Graph Embedding}
 
	Inspired by GraphSAGE\cite{hamilton2017inductive}, we propose a novel graph embedding method using a shifted feature aggregation approach to achieve graph representations. As shown in Fig.~\ref{fig:graph_embedding}(c), feature representations of each satellite are divided into three areas: current satellite feature, aggregated feature of 1-hop neighbors, and aggregated feature of 2-hop neighbors. As shown in Fig.~\ref{fig:graph_embedding}(a), in the feature aggregating mechanism, each satellite receives the graph representations of its neighbors, aggregates the features of the 1st and 2nd areas of its neighbors, shifts them to the 2nd and 3rd areas, and concatenate them with the feature representation of current node to form a new graph representation. As shown in Fig.~\ref{fig:graph_embedding}(b), the features of the neighbors along with the feature of the current node provide information within a 3-hop range. The differences in features in different areas among neighbors enable the extraction of precise directional information and resource distribution, which is highly beneficial for the computing decision and routing selection. Additionally, this graph embedding mechanism naturally limits the feature propagation distance to 3 hops, eliminating the non-timely information from distant nodes.
	
	In GAT, the algorithm calculates weights for each neighbor and normalizes them using the Softmax function\cite{velivckovic2017graph}. However, when precise node-level features are needed, this method can overestimate or underestimate information about certain nodes. To achieve a more stable aggregation, we use an equal weight summation method for feature aggregation, setting the weight to $\frac{1}{K}$ where $K$ is the maximum degree in the graph (i.e., the largest number of neighbors of any node in the graph). Assuming the current node has $I$ neighbors, with the graph representation of $j^{th}$ neighbor being $h_{j}=[h^1_{j},h^2_{j},h^3_{j}]^{T}$, where $h^1_{j}$, $h^2_{j}$, and $h^3_{j}$ are the 1st area, 2nd area and 3rd area in the graph representation of $j^{th}$ neighbor, respectively. The shifted feature representation of this neighbor is $h'_{j}=[O,h^1_{j},h^2_{j}]^{T}$, where $O$ is a zero vector. If the feature of the current node is $h^1_{v_{i}}$, and we use zero vectors for filling areas 2 and 3 $h'_{v_{i}}=[h^1_{v_{i}}, O, O]^{T}$, then the aggregated feature of the current node is:
	\begin{equation}
		h^*_{v_{i}}=h'_{v_{i}}+\frac{1}{K}\sum_{j=1}^{I} h'_{j}
	\end{equation}
    where $+$ represents element-size addition.
 
	Additionally, when aggregating 2-hop neighbor information through features of 1-hop neighbors of neighboring satellites, subtracting the self-node feature part from neighbor features during aggregation can decouple different areas in aggregated features, which is very helpful for the training of DRL readout. Thus, let $h''_{j}=[O,h^1_{j},\frac{K}{K-1} h^2_{j}]^{T}$, $h''_{v_{i}}=[h^1_{v_{i}},O,-\frac{I}{K(K-1)}h^1_{v_{i}}]^{T}$, the corrected aggregated graph representation of current node:
	\begin{equation}
		\begin{split}
			h_{v_{i}} = h''_{v_{i}} + \frac{1}{K} \sum_{j=1}^{I} h''_{j}
		\end{split}
	\end{equation}
	
	When analyzing 2nd area and 3rd area independently, we obtain the following relationships:
	\begin{equation}
		h^{2}_{v_{i}}=\frac{1}{K}\sum_{j=1}^{I} h^{1}_{j}
	\end{equation}
	
	\begin{equation}
		h^{3}_{v_{i}}=\frac{1}{K-1}\sum_{j=1}^{I} h^{2}_{j} -\frac{I}{K(K-1)}h^1_{v_{i}} 
	\end{equation}

\subsubsection{Initial Padding and Fault Padding}

Since each feature aggregation requires the embedding information of neighboring nodes, it is necessary to fill the positions of neighbors that have not yet generated graph embeddings during system initialization. According to section \ref{sec:representation}, the graph representation of an idle satellite $v_{0}$ with idle neighbors can be represented as $h_{0}=\{0,1,0,0,0,1,0,0,0,1,0,0\}$.

In addition, when neighboring satellites experience faults, special padding is required to mark them, so that the fault information can be represented during feature aggregation. we use the opposite features $h_{j}=\{1,0,2 \cdot h^a_{3},2 \cdot h^a_{4},1,0,2 \cdot h^a_{7},2 \cdot h^a_{8},1,0,2 \cdot h^a_{11},2 \cdot h^a_{12}\}$ to fill in for unavailable satellites. Where $h^a_{k}$ donates The average value of the features of other neighbor graph embeddings at the $k$th position. The idea is to mark the remaining storage space in the direction of the satellite as 0, treating it as unreachable. Meanwhile, the occupation of computing and transmission resources is overestimated by using twice the average value of the resources of the other neighbors. Thus, the occupation of computing and transmission resources will be estimated as $K/I$ times after feature aggregation.

\section{DRL-based Solution for Joint Optimization}
\label{sec:drl}

A DRL algorithm is proposed to address the issue of distributed decision-making for joint computing and routing in complex satellite networks. Given the inherent limitations in obtaining complete system information in satellite networks, the POMDP framework is employed for dynamic optimization under conditions of partial visibility.

\subsection{POMDP}

\subsubsection{State Space}

When a task arrives, the task information $S_{t}$ and local resource information $S_{r}$ along with the destination state $S_{d}$ will serve as input state to the DRL agent, i.e., $S=\{S_{r},S_{t},S_{d}\}$.

\begin{itemize}
\item The resource state $S_{r}$ can be represented by resource information of the current satellite, received graph representations of neighbors, and edge information, i.e., $S_{r}=\{h^1_{i},h_{1},...,h_{K}, E\}$, in an LEO satellite constellation, $K=4$. To address the requirement of neural networks for a fixed input dimension, specific encoding is employed to supplement missing neighbors when the number of neighbors falls short of $K$. The edge information, which includes the queuing information of transmission for ISL towards each neighbor satellite, is critical for optimizing task delay. $E=\{q^{t}_{i,j},...,q^{t}_{i,K}\}$, where $q^{t}_{i,j}$ represents the transmission queue length towards the $j^{th}$ neighbor, normalized by the total storage capacity of the satellite. When the number of neighbors is less than $K$, the missing positions are filled with a value of 1, signifying that the link in that direction is unreachable.
\item The task state $S_{t}$ can be formed by a direction vector of destination $S^{D}_{t}$ and current task state $S^{s}_{t}$, i.e., $S_{t}=\{S^{D}_{t},S^{s}_{t}\}$. $S^{D}_{t}=\{l_{1},...,l_{K}\}$, where $l_{j}$ represents the number of hops from the $j^{th}$ neighbor to the destination satellite, normalized by dividing by the diameter of the network graph. If the number of neighbors is less than $K$, the gaps are filled with a value of 2, indicating that the link in that direction is unreachable. Task state $S^{s}_{t}=\{s,d,s',x_{c}\}$, where $s$ is the data size of the task, $d$ is computation demand, $s'$ is post-computation data size, and $x_{c}$ is a binary variable indicating whether the task has been computed.
\end{itemize}
\subsubsection{Action Space}
The action space can be represented as $\mathbf{A} = \{A_{1},..., A_{K}, A_{c}\}$, where $A_{j}$ denotes pushing the task to the transmission queue towards the $j^{th}$ neighbor, and $A_{c}$ denotes pushing the task to the computing queue of the current satellite.

\subsubsection{Reward}
The reward function $R$ for task allocation in satellite networks is defined based on various conditions to optimize task efficacy while minimizing delays and resource usage. Each condition is associated with a specific transmission or processing event, with corresponding rewards or penalties:

\begin{enumerate}
	\item When the task reaches the destination ground station, the reward is calculated as:
\begin{equation}
	R =
	\begin{cases}
		\beta_{s} + \beta_{d} \cdot (t_{b} - t_{\tau}), & \text{if }x_{c} = 1 \\
		\beta_{d} \cdot (t_{b} - t_{\tau}), & \text{if } x_{c} = 0
	\end{cases}
\end{equation}
where $x_{c}$ donates whether the task is computed. $\beta_{s}$ is the reward for successful transmission, and $\beta_{d}$ is the delay penalty. Here, $t_{b}$ is the task start time, and $t_{\tau}$ is the current decision time.

\item When packet loss occurs during transmission. The reward is:
\begin{equation}
	R = -\beta_{l} + \beta_{d} \cdot (t_{b} - t_{\tau}),
\end{equation}
where $\beta_{l}$ is the penalty for packet loss.

\item For a normal one-step transition:
\begin{equation}
	R = 
	\begin{cases}
		\beta_{d} \cdot (t_{L} - t_{\tau}), & \text{if } m_{i} < m_{r} \\
		-\beta_{m} + \beta_{d} \cdot (t_{b} - t_{\tau}), & \text{if } m_{i} >= m_{r}
	\end{cases}
\end{equation}
where $m_{r}$ is the reserved storage space. $t_{L}$ representing the time at the last decision step, $\beta_{m}$ is the penalty for exceeding the storage threshold.
	
\end{enumerate}

These conditions collectively aim to ensure that tasks are processed and transmitted efficiently while managing the limited resources of satellite networks effectively. In our experiment, $\beta_{s}$, $\beta_{d}$, $\beta_{l}$, and $\beta_{m}$ are 1, 0.05, 1, and 0.25, respectively.

\subsection{D3QN}

\begin{figure}[!t]
	\centering
	\includegraphics[width=0.475\textwidth]{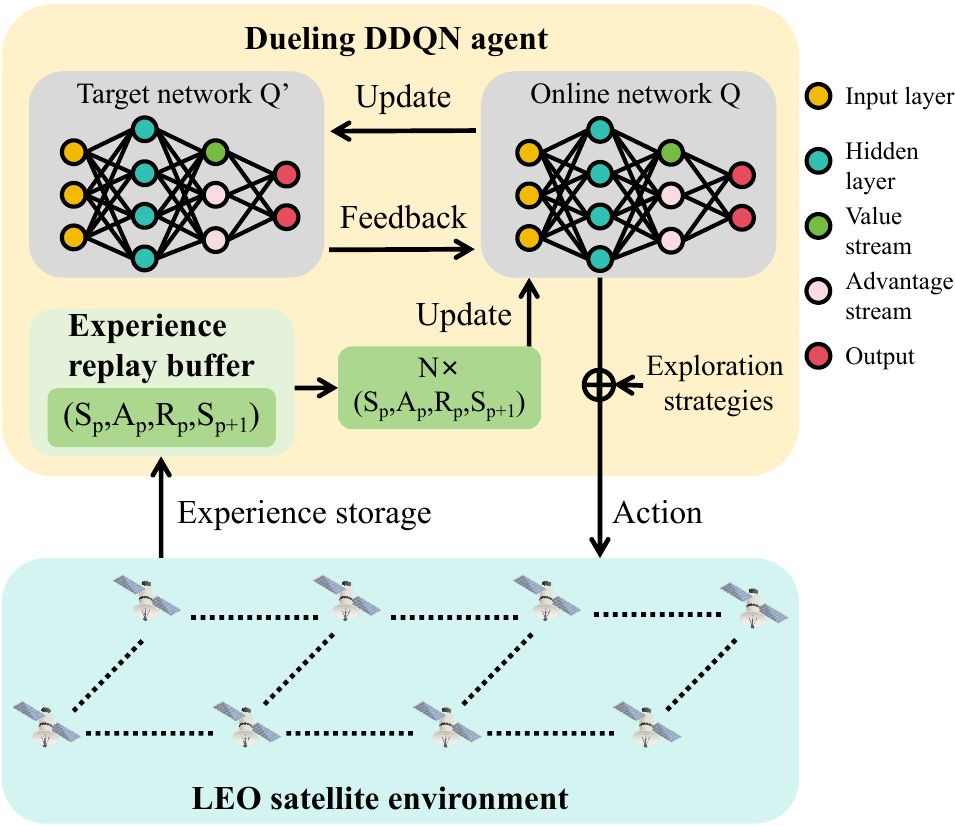}
	\caption{D3QN-based joint optimization of computing and routing for LEO satellites in training.}
	\label{fig2}
\end{figure}

\textit{Double deep Q-Network} (DDQN) \cite{van2016deep} is a notable deep reinforcement learning technique designed to address the overestimation issue inherent in traditional Q-learning algorithms by introducing dual Q-networks. The dueling architecture further improves this approach by decomposing the Q-value into separate estimates of the state-value and advantage functions, facilitating more robust policy learning and enhanced performance\cite{wang2016dueling}.

As shown in Fig.~\ref{fig2}, the core framework of D3QN comprises two main components: an online network $Q$ for selecting optimal actions and a target network $Q'$ for evaluating these actions. Both networks have the same structure, consisting of some hidden layers to extract features from the state $S$, followed by two streams: one for estimating the state-value function $V(S)$ and the other for estimating the advantage function $\mathcal{A}(S, A)$. The Q-value is then computed by combining these two streams as follows:

\begin{equation}
	Q(S, A) = V(S) + \left( \mathcal{A}(S, A) - \frac{1}{|\mathbf{A}|} \sum_{A' \in \mathcal{A}} \mathcal{A}(S, A') \right)
\end{equation}
where $|\mathbf{A}|$ denotes the number of possible actions.

The online network $Q$ selects actions $A$ based on the current state of the environment $S$ and updates its parameters at each step to quickly adapt to environmental changes. In contrast, the target network $Q'$ updates its parameters less frequently to maintain stability in the learning process.

The target value $y$ is calculated using the target network $Q'$. For each experience tuple $(S, A, R, S')$, the target value $y$ is defined as the current reward $R(S, A)$ plus the product of the discount factor $\gamma$ and the predicted value of target network $Q'$ for the next state $S'$ and best next action $A'$. Specifically, $\gamma$ is a value between 0 and 1 that determines the degree to which the algorithm prioritizes short-term versus long-term rewards. A lower value of $\gamma$ leads the agent to emphasize immediate rewards, while a higher value encourages the agent to prioritize long-term rewards. This can be expressed as:

\begin{equation}
	y=R(S,A)+\gamma \cdot \max_{A' \in \mathbf{A}}(Q'(S',A',\theta'))
\end{equation}
where $\theta'$ represents the weights of target network $Q'$.

The loss function $L(\theta)$ is based on the temporal difference loss, measuring the difference between the predicted value of the online network and the target value. Here, $\theta$ represents the weights of the online network $Q$. It is given by:

\begin{equation}
L(\theta) = [y-Q(S,A,\theta)]^{2}
\end{equation}
Minimizing this loss function enables D3QN to effectively learn the optimal policy.

The parameters of the online network are updated using gradient descent to minimize the loss function. The formula for updating is:

\begin{equation}
	\theta = \theta - \alpha \cdot \nabla_{\theta} L(\theta)
\end{equation}
where $\alpha$ is the learning rate, and $\nabla_{\theta} L(\theta)$ represents the gradient of the loss function with respect to the parameters of the online network. During the update process, the algorithm computes the gradient of the loss function concerning the parameters of the online network and adjusts them accordingly to optimize the action selection.

\begin{algorithm}[!t]
	\caption{D3QN for joint optimization of computing and routing}
	\label{alg:DDQN}
	\begin{algorithmic}[1]
		\STATE \textit{Initialization}:
		\\ Initialize $Q(\theta)$ with random weights and copy them to $Q'(\theta')$. Establish replay buffer $B$, set initial exploration probability $\epsilon$ and heuristic exploration probability $P_{h}$.
		\FOR {each training epoch}
		\STATE Reset the initial state $S_{0}$ and set marker $p=0$
		\FOR {each task within the epoch}
		\WHILE {task not completed or lost}
		\IF {a sample from [0,1] \textless $\epsilon$}
            \IF {a sample from [0,1] \textless $P_{h}$}
            \STATE Select $A_{p}$ using heuristic exploration strategy
            \ELSE
		\STATE Select $A_{p}$ randomly
            \ENDIF
		\ELSE
		\STATE Select $A_{p}$ that maximizes $Q(S_{p},A_{p},\theta)$
		\ENDIF
		\STATE Execute $A_{p}$, observe $S_{p+1}$ and $R_{p}$
		\STATE Store $(S_{p}, A_{p}, R_{p} S_{p+1})$ in the replay buffer $B$
		\STATE $p \leftarrow p+1$
		\ENDWHILE
		\ENDFOR
		\FOR {each training iteration}
		\STATE Sample a batch of experiences from $B$
		\STATE $y=R(S,A)+\gamma \cdot \max_{A' \in \mathbf{A}}(Q'(S',A',\theta'))$
		\STATE Calculate $L(\theta) = [y-(Q(S,A,\theta))]^{2}$
		\STATE Update weights of network $\theta \leftarrow \theta - \nabla_{\theta} L(\theta)$
		\ENDFOR
		\STATE Update exploration probability $\epsilon$
		\IF {current epoch is a multiple of the target update period}
		\STATE Copy parameters from the online network $Q(\theta)$ to the target network $Q'(\theta')$
		\ENDIF
		\ENDFOR
		\STATE Store the parameters of $Q(\theta)$
	\end{algorithmic}
\end{algorithm}

\subsection{Some tricks in D3QN}
\subsubsection{Improvement of Action Selection}
In standard D3QN networks, action selection is typically predicated on the action with the highest Q-value in the current state. In our improved version, which is tailored for computing and routing integrated tasks, this process is modified to ensure better suitability. Specifically, when the task has been computed ($x_{c}=1$), the agent selects the action with the highest Q-value only from the transmission actions, i.e., $\mathbf{A^{*}} = \{A_{1},..., A_{K}\}$. When the task state is 0 (the task not yet computed), the agent selects the action with the highest Q-value from all possible actions, i.e., $\mathbf{A} = \{A_{1},..., A_{K}, A_{c}\}$.

Similarly, when the target network $Q'$ evaluates Q-values for the next state $S'$, it chooses the highest Q-value from different action sets based on the computation state of the next task. This improvement allows the DRL agent to assess the most valuable action in a given state more precisely, thereby improving overall decision quality:
\begin{equation}
    \begin{aligned}
        y = \left\{
        \begin{aligned} 
            &R(S,A)+\gamma \cdot \max_{A' \in \mathbf{A}}(Q'(S',A',\theta')), &&\text{if $x'_{c} = 0$}, \\
            &R(S,A)+\gamma \cdot \max_{A' \in \mathbf{A^{*}}}(Q'(S',A',\theta')), &&\text{if $x'_{c} = 1$}.
        \end{aligned}
        \right.
    \end{aligned}
\end{equation}

\subsubsection{Heuristic Exploration Strategy}
The original $\epsilon$-greedy strategy employs a completely random approach during exploration, which may lead to a long convergence period. Therefore, in the exploration strategy, a heuristic exploration method is introduced and used in conjunction with the purely random approach. In the heuristic exploration method, the computing action $A_{c}$ is selected with a certain probability when the task has been computed ($x_{c}=1$). In the case of choosing transmission actions, the next hop for transmission is selected based on the shortest number of hops to the destination. During training, this exploration strategy can improve the efficiency of learning and accelerate convergence.

This combination of an improvement of action selection and heuristic exploration strategy enables the agent to handle tasks more flexibly and effectively.

\subsection{Complexity Analysis}
\label{sec:complexity}
\subsubsection{Complexity Comparison between Centralized and Distributed Approaches}
According to\cite{cao2023computing}, in a centralized strategy, the decision-making process can be divided into three parts: selection of the computing satellite, routing selection from the source node to the computing satellite, and routing selection from the computing satellite to the destination satellite. Assuming the network has $N$ nodes, for each candidate computing node, planning of two routing paths is required: from the source satellite to the computing satellite and from the computing satellite to the destination. The shortest path can be calculated using a shortest path algorithm such as Dijkstra, with a computational complexity of $O(N \cdot log_{2}\sqrt{N})$ in binary heap. Since the number of candidate computing nodes is proportional to the total number of network nodes, the time complexity of a single decision is $O(N^{2} \cdot log_{2}\sqrt{N})$. Therefore, as the scale of the network increases, the computational complexity of a decision increases rapidly. 

As for the proposed distributed method, assuming a DRL agent with a hidden dimension of $H$, the single-step time complexity is about $O(H^{2})$. The average distance between two nodes is proportional to the diameter of the network, i.e., the average hop count of transmission is proportional to $\sqrt{N}$, making the cumulative computational complexity of multi-step decisions $O(H^{2} \cdot \sqrt{N})$. Usually, $H$ can be a fixed number at various constellation scales. Therefore, although the distributed strategy increases decision times, its time complexity could be significantly lower compared to the centralized solution when a large constellation is considered.

\subsection{Complexity Comparison of Different Graph Methods}

In graph neural network methods such as MPNN\cite{gilmer2017neural} and GAT\cite{velivckovic2017graph}, each node in the graph aggregates features from its immediate neighbors, which typically involves computations like weighted sum (in MPNN) or attention-weighted sum (in GAT). In the readout stage, features of all nodes in the graph are necessary. This means that in a satellite network with $N$ satellites, both the time and space complexities of the graph neural network will be $O(N)$. Furthermore, distributed implementations of graph neural network methods\cite{Zhou2025Distributed}\cite{He2025Routing} face even greater challenges, as each node at a distributed approach must assume the role of message aggregation, leading to time and space complexities of $O(N^2)$ at the network-scale.
	
Similar to GraphSAGE\cite{hamilton2017inductive}, the graph embedding method proposed in this paper samples neighboring nodes within subgraphs and extracts the graph representation from the features of certain nodes. Assuming the average number of neighboring nodes is $K$, the time and space complexities are $O(K)$ per node. So that complexities are $O(K \cdot N)$ across the entire network. Since $K$ is much smaller than $N$, the proposed method reduces the computational time complexity and communication overhead significantly compared to other existing methods.

\section{Simulation Results and Analysis}
\label{sec:result}
\subsection{Experiment Setup}

\begin{table}[!t]
    \caption{Experimental Parameter Settings\label{tab:sim-config}}
      \scriptsize
    \begin{tabular}{p{4.75cm}p{3.25cm}}
        \hline
        \noalign{\vspace{1.5pt}}
        PARAMETERS & VALUES\\\hline
        \noalign{\vspace{1.5pt}}
            \; \footnotesize \textbf{Environment Parameters}\\
        Number of orbital planes & 12 \\
        Number of satellites per plane & 24 \\
        Orbital altitude & 500 km \\
        Orbital inclination & 60° \\
        Beam coverage angle & 45° \\
        Data size per task & 25-75 MB \\
        Compression ratio& 9-11 \\
        Computation demand for compression & 1200-2000 FLOP/Byte \\
        The data size of inference output & 5 KB \\
        Computation demand for inference & 2400-4000 FLOP/Byte \\
        Generation frequency of task & 1 times/s \\
        Average observation duration & 40s \\
        Update frequency of graph representation & 10 times/s \\
        Detection period of link failure & 0.25s \\
        ISL transmission rate & 1.2 Gbps \\
        Downlink transmission rate & 3 Gbps \\
        Onboard computing capacity & 50 GFLOPS \\
        Satellite storage limit & 1 GB \\
        \hline \noalign{\vspace{1.5pt}}
           \; \footnotesize \textbf{DRL Parameters}\\
        Optimizer & Adam (non-Amsgrad mode) \\
        Learning rate ($\alpha$) & 0.0002 \\
        TD target decay ($\gamma$) & 0.99 \\
        Number of hidden layers & 2 \\
        Number of neurons in hidden layers & 256 \\
        Activation function & LeakyRelu (slope = -0.01) \\
        Number of training epochs & 6000 \\
        Size of experience replay pool  & 200000 \\
        Size of mini-batch & 1024 \\
        Initial exploration probability & 0.9 \\
        Heuristic exploration probability & 0.5 \\
        Exploration probability decay & 0.999 \\
        Minimum exploration probability & 0.02 \\
        \hline
    \end{tabular}
\end{table}


To evaluate the performance of different algorithms in the LEO satellite constellation, a system-level simulator is developed in \textit{Python}. The discrete event library \textit{SimPy} is used to build computing and transmission queues. Library \textit{skyfield} is used for satellite ephemeris calculations. The DRL network was constructed using the \textit{PyTorch} framework. By setting the same random seed, identical tasks, and link failures are generated across different environments, ensuring the evaluation of various algorithms under the same high-dynamic scenarios. 
The detailed experimental parameter settings are shown in Table \ref{tab:sim-config}. Furthermore, in order to analyze the generalization ability of the algorithm on different-sized constellations, this paper conducted tests on four different-sized constellations: Iridium, Telesat, OneWeb, and Starlink (Gen1 shell 2). The detailed parameters of each constellation are shown in Table \ref{tab:constellation_params}. Ground stations were distributed according to Table \ref{tab:geo-coords}. A connection between satellites and ground stations can only be established when the ground station is located within the coverage area of the satellite beam. When random tasks were generated, the nearest satellite connected to the ground station was selected as the destination for data backhaul. Owing to the exceedingly high computation cost and runtime in constellation-scale simulation, a low storage resource limit (1GB) was adopted. This approach was employed to evaluate the storage resource planning capability of the algorithm in a short simulation time.All experiments are conducted on a workstation equipped with an NVIDIA RTX 4080 Ti GPU (16~GB), a 14-core CPU, and 32~GB of RAM. 

\subsection{Service Model}

On each observation satellite, task arrivals follow a Poisson process with rate $\lambda$, and service times follow an exponential distribution with rate $\mu$, forming a standard M/M/1 queue model. This modeling captures random and independent arrivals and services with constant average rates. Considering that ground observation demand is higher than ocean observation demand, the task arrival interval for satellites covering ocean regions is set to be twice that of satellites covering ground regions. Moreover, satellites with direct connections to ground stations transmit local observation data directly to the ground without onboard processing.

\subsection{Link Failure Model}

\begin{table}[!t]
\centering
\caption{Satellite Constellation Parameters\label{tab:constellation_params}}
\scriptsize
\setlength{\tabcolsep}{8pt}        
\renewcommand{\arraystretch}{1.3} 
\setlength{\arrayrulewidth}{0.3pt} 

\begin{tabular}{|c|c|c|c|c|}
\hline
\multirow{2}{*}{\textbf{PARAMETERS}} &
\multicolumn{4}{c|}{\textbf{Constellation Name}} \\ \cline{2-5}
& \textbf{Iridium} & \textbf{Telesat} & \textbf{ OneWeb} & \textbf{Starlink} \\
\hline
\textbf{\textit{Orbital Altitude}}  & 780 km & 1015 km & 1200 km & 570 km  \\ \hline
\textbf{\textit{Orbital Inclination}} & 86.4° & 98.98° & 87.9° & 70°  \\ \hline
\textbf{\textit{Orbital Planes}}& 6 & 27 & 18 & 36 \\ \hline
\textbf{\textit{Satellites per Plane}} & 11 & 13 & 36 & 20  \\ \hline

\end{tabular}
\end{table}

\begin{table}[!t]
	\centering
	\caption{Locations of Ground Stations in Simulation\label{tab:geo-coords}}
	\begin{tabular}{| c | c | c |}
		\hline
		\textbf{City} & \textbf{Latitude (\textdegree)} & \textbf{Longitude (\textdegree)} \\ \hline
		Dubai & 25.252 & 55.280 \\
		Harbin & 45.750 & 126.650 \\
		Istanbul & 41.019 & 28.965 \\
		Jakarta & -6.174 & 106.829 \\
		Karachi & 24.867 & 67.050 \\
		Moscow & 55.752 & 37.616 \\
		Nairobi & -1.283 & 36.817 \\
		Sanya & 18.243 & 109.505 \\
		Shanghai & 31.109 & 121.368 \\
		Urumqi & 43.800 & 87.583 \\
		Xian & 34.258 & 108.929 \\
		\hline
	\end{tabular}
\end{table}

The link failure model can be described as a Markov process where each satellite link has two possible states: normal and faulty. At each time step, the link has a certain probability $p$ to transition from normal to faulty. Once in the faulty state, the link has a recovery probability $q$ to transition back to normal. The transition probabilities form a two-state Markov chain with the transition matrix:

\begin{equation} 
P =
\begin{pmatrix}
1 - p & p \\
q & 1 - q
\end{pmatrix}
\end{equation}
where $p$ represents the probability of failure, and $q$ is the probability of recovery.

The average failure probability $\pi_{1}$ of the satellite link can be calculated as the steady-state probability of the link being in the faulty state. This is given by the formula:

\begin{equation} 
\pi_{1}= \frac{p}{p + q}
\end{equation}

\subsection{Comparing Algorithms}
\begin{figure}[t]
    \centering
     \includegraphics[trim=0pt 10pt 0pt 5pt, clip, width=0.475\textwidth]{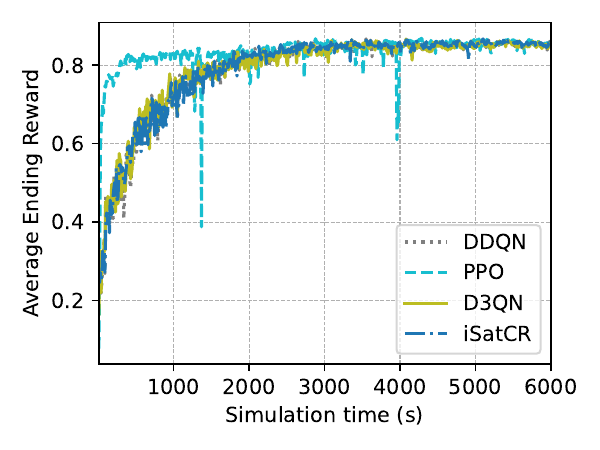}
    \caption{Average reward changing curve during DRL training. }
    \label{fig: reward variation}
\end{figure}

To provide a comprehensive evaluation of its performance, the proposed iSatCR method is benchmarked against a diverse set of four baseline algorithms. 
\begin{itemize}
    \item Dueling Double Deep Q-Network (D3QN): This method uses observations of neighboring satellites as state inputs, without incorporating graph embeddings.
    \item Double deep Q-network (DDQN): This method uses neighboring satellite information as input state. Unlike D3QN and iSatCR, it does not utilize dueling networks.
    \item Proximal policy optimization (PPO)\cite{schulman2017ppo}: A DRL algorithm based on the Actor-Critic architecture and policy optimization methods. It improves performance by iteratively updating the policy and value function using a clipped objective function to maintain stable learning, prevent large policy updates, and ensure efficient exploration and exploitation.
    \item Ideal centralized solution (ICS): The centralized approach 
    adopts the algorithm proposed in\cite{cao2023computing}, which plans routing based on global information. The original algorithm was found to be inadequate for adapting to the complex and highly dynamic topology changes of the simulation environment which arise from random failures and unforeseen traffic contingencies. Consequently, we integrate it with the resource reservation mechanism proposed in this paper, and its heuristic algorithm, which was based on genetic algorithms, has been replaced by an optimal algorithm that searches for all feasible solutions. 
\end{itemize}

In our experiment, DRL methods(iSatCR, D3QN, DDQN, and PPO) share identical configurations. These DRL methods are distributed, making decisions at each step of task execution, and all decision-making processes within the simulation took into account the latency effects, which included both the update frequency and the propagation delay. The global information used for decision-making in a centralized solution was obtained instantly from the simulator, ignoring all latencies in information updates. 

\subsection{Results Analysis}

\subsubsection{Rewards in DRL Training}

Fig.~\ref{fig: reward variation} illustrates the convergence performance of multiple algorithms during the training phase. The PPO algorithm, which benefits from action sampling for exploration, converges fastest due to its effective balance between exploration and exploitation. However, its performance after convergence shows instability, which is evident from frequent and significant fluctuations in the reward curve. In contrast, DDQN, D3QN, and iSatCR algorithms, which utilize the $\epsilon$-greedy method, converge more slowly due to enforced random exploration in the early training phase. Nevertheless, these algorithms achieve greater stability after convergence, with minimal fluctuations in their reward curves. After convergence, all algorithms exhibit comparable performance, with iSatCR achieving slightly higher rewards than the others.

\subsubsection{Simulation Results in Different Task Load}
\begin{figure*}[ht]
    \centering
        \subfloat[]{
        \includegraphics[width=0.315\textwidth]{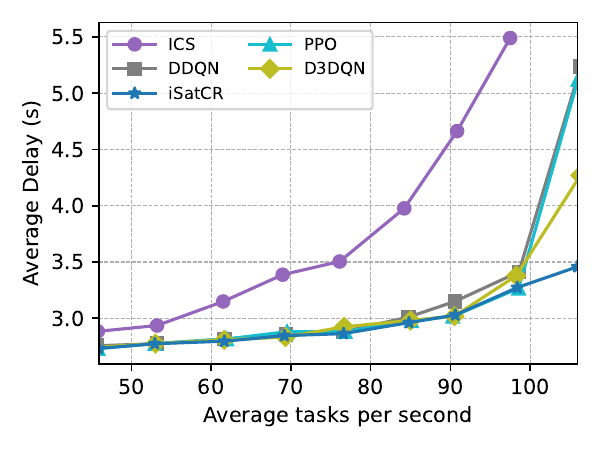}%
        \label{fig:delay_dif_load}
        }
    \hfill
        \subfloat[]{
        \includegraphics[width=0.315\textwidth]{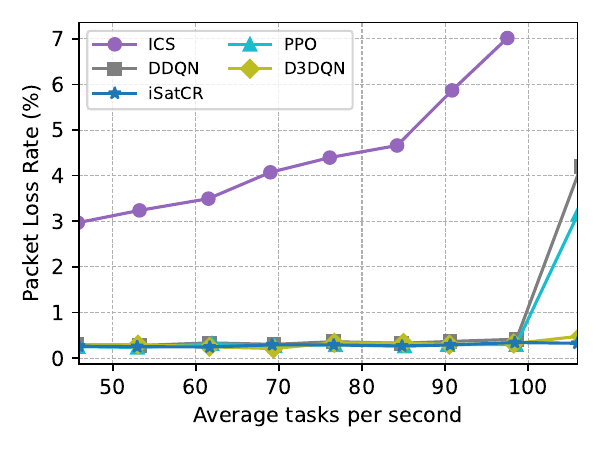}%
        \label{fig:loss_dif_load}
        }
    \hfill
        \subfloat[]{
        \includegraphics[width=0.315\textwidth]{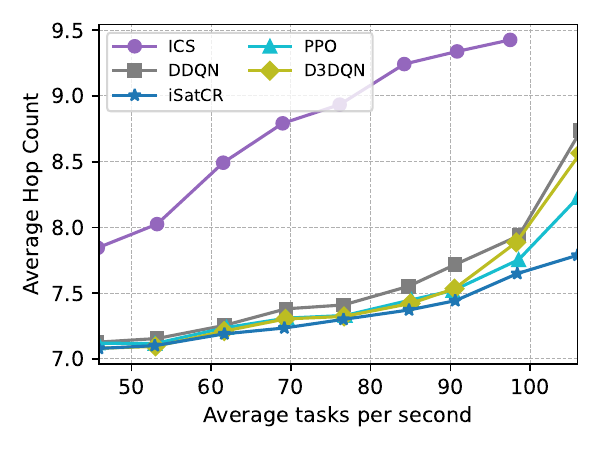}%
        \label{fig:hops_dif_load}
        }
    \caption{Simulation results of the satellite constellation under various algorithms and traffic loads. (a) Comparison of average task delay, including transmission, propagation, computation, and queuing delays. (b) Comparison of packet loss rates. (c) Comparison of average transmission hop count.}
    \label{fig:dif_load}
\end{figure*}

We fixed the link failure rate at 3\% and demonstrated the adaptability of the iSatCR algorithm to different loads by presenting the average task latency, average packet loss rate, and average number of transmission hops of all algorithms under varying task loads. Fig.~\ref{fig:delay_dif_load} illustrates the relationship between the average task delay and task load under different algorithms.At lower task loads, the ICS and DRL algorithms perform similarly. However, as the task load increases, the centralized method ICS experiences a rapid rise in task delay, while DRL methods experience less pronounced delay increases. This highlights the superior adaptability of distributed DRL algorithms, which can make decisions based on real-time network conditions dynamically, particularly in high-load satellite networks. For task rates under 85 tasks per second, the three DRL algorithms deliver comparable performance. Under higher task loads, DDQN, PPO, and D3QN algorithms exhibit a sharp increase in delay. In contrast, the graph embedding-based algorithm iSatCR maintains lower delays and achieves the best performance across all load conditions. These results indicate that under low loads, one-hop neighbor information is enough for task planning. Graph embedding enables agents to consider multi-hop neighbor information, ensuring low delays even under high-load conditions.

Fig.~\ref{fig:loss_dif_load} depicts the variation in average packet loss rate with task load across different algorithms. Four DRL algorithms demonstrate robust adaptability to dynamic network conditions, maintaining packet loss rates below 1\% in most traffic loads. In contrast, the ICS algorithm exhibits significantly higher packet loss rates, which rise sharply with increasing load, ranging from 3\% to 7\%. This illustrates the limitations of centralized methods in handling sudden link failures and traffic bursts. When the task load at 105 tasks per second, DDQN, D3QN, and PPO algorithms show different increases in packet loss rate. At the same time, iSatCR maintains a low packet loss rate even in high load conditions.

The relation between average transmission hops and task load for different algorithms is presented in Fig.~\ref{fig:hops_dif_load}. The trend in average hops is similar to that of task delay, suggesting that increased delay is largely driven by the selection of more distant computing satellites and longer transmission paths. Among the algorithms, iSatCR achieves the smallest average number of hops, showcasing its ability to identify closer computing satellites for task execution. Its capability effectively minimizes the number of processing and transmission steps, improving overall performance.

\subsubsection{Simulation Results in Different Link Failure Rate}

\begin{figure*}[ht]
    \centering
        \subfloat[]{
        \includegraphics[width=0.315\textwidth]{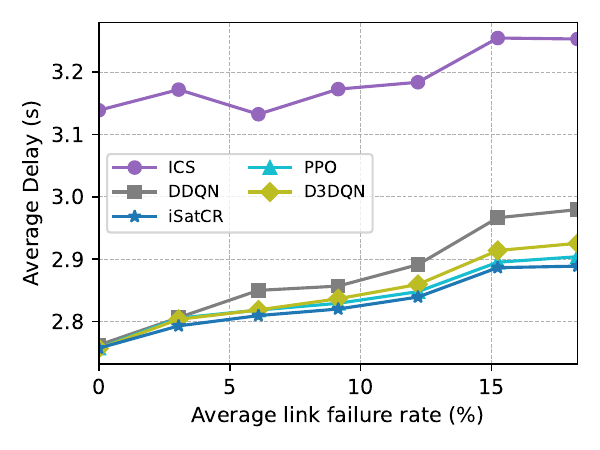}%
        \label{fig:delay_dif_fail}
        }
    \hfill
        \subfloat[]{
        \includegraphics[width=0.315\textwidth]{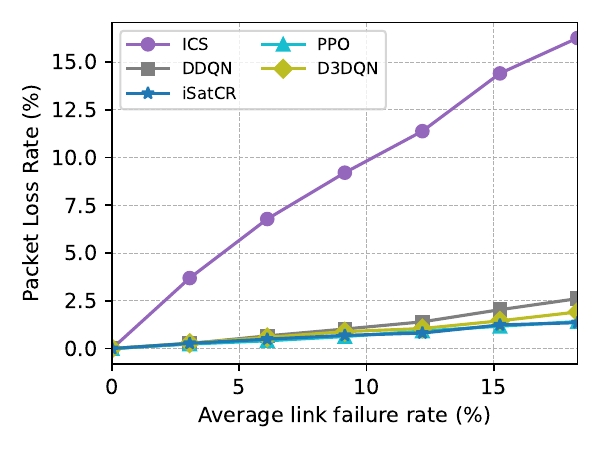}%
        \label{fig:loss_dif_fail}
        }
    \hfill
        \subfloat[]{
        \includegraphics[width=0.315\textwidth]{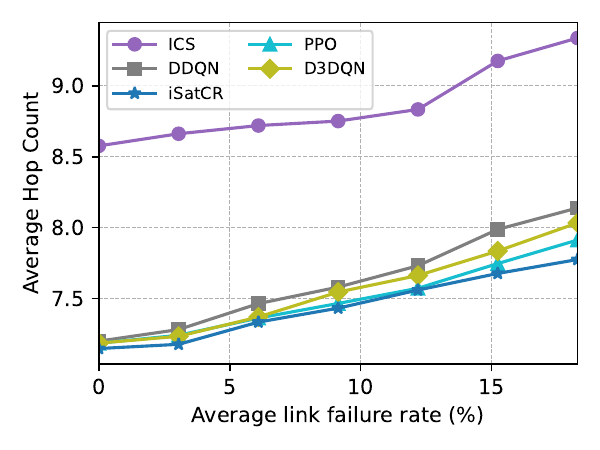}%
        \label{fig:hops_dif_fail}
        }
    \caption{Simulation results of the satellite constellation under various algorithms and link failure rate. (a) Comparison of average task delay. (b) Comparison of packet loss rates. (c) Comparison of average transmission hop count.}
    \label{fig:dif_fail}
\end{figure*}

To evaluate the robustness of the iSatCR algorithm, we fixed the task load at 70 tasks per second and tested the performance of all algorithms under different link failure rates. Fig.~\ref{fig:delay_dif_fail} shows the variation of the average task delay of each algorithm with the link failure rates. Among the three DRL algorithms, DDQN experiences the fastest increase in delay as the link failure rate rises. However, with the graph embedding method, iSatCR maintains the lowest delay across all varying link failure conditions, highlighting its robustness in performance.

As depicted in Fig.~\ref{fig:loss_dif_fail}, the variation in average packet loss rates with link failure rates reveals significant differences in algorithm performance. In scenarios without link failures, all algorithms maintain zero packet loss. However, with increasing link failure rates, the ICS method exhibits a rapid rise in packet loss, demonstrating its limited capability to handle sudden link disruptions. Among the DRL algorithms, DDQN shows the worst performance, whereas iSatCR achieves the lowest packet loss rates. These results indicate that the extended awareness range provided by the graph embedding method effectively reduces packet loss.

Fig.~\ref{fig:hops_dif_fail} presents the variation in average transmission hops as link failure rates increase. Across all algorithms, the average transmission hops rise due to the inaccessibility of optimal paths, necessitating the selection of longer alternative routes. Notably, iSatCR exhibits the smallest hop increase, underscoring its superior decision-making and route-planning capabilities in high-failure scenarios.

\subsubsection{Distribution of Satellite Computing Time}
\begin{figure}[t]
    \centering
    \includegraphics[trim=0pt 7pt 0pt 5pt, clip, width=0.475\textwidth]{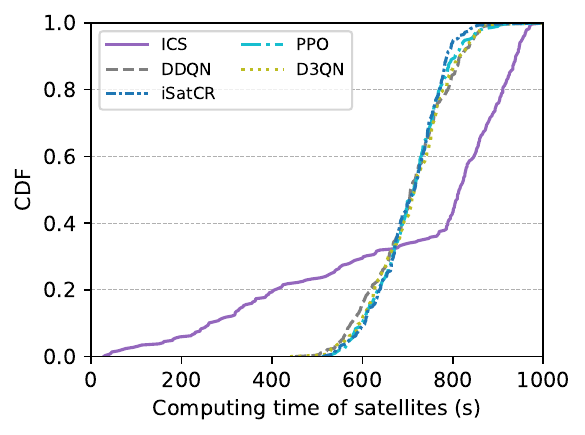}
    \caption{Cumulative distribution of satellite computing time under different algorithms}
    \label{fig:cdf_20}
\end{figure}
Fig.~\ref{fig:cdf_20} illustrates the cumulative distribution of satellite computing time at a task load of 90 tasks per second, highlighting the load-balancing capabilities of various algorithms. A steeper curve signifies a more concentrated distribution of computing time, corresponding to improved load balancing. The ICS algorithm shows a broader distribution, with both low-load and high-load satellites prominently represented. In contrast, the DRL algorithms exhibit more balanced distributions. Among these, iSatCR achieves the steepest cumulative distribution curve, indicating the most balanced computing load among all algorithms.

\subsubsection{Simulation Results in Different-Scale Constellations}

\begin{figure*}[ht]
    \centering
        \subfloat[]{
        \includegraphics[width=0.315\textwidth]{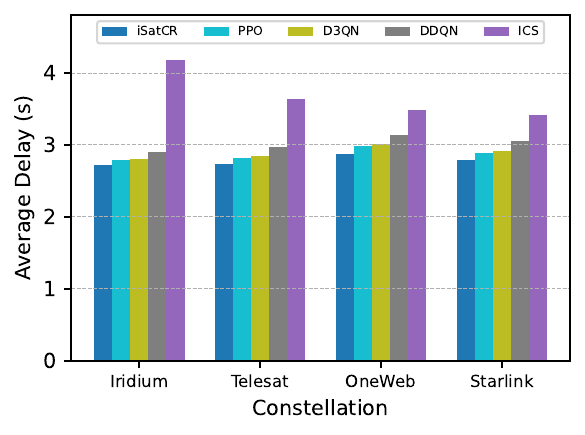}%
        \label{fig:delay}
        }
    \hfill
        \subfloat[]{
        \includegraphics[width=0.315\textwidth]{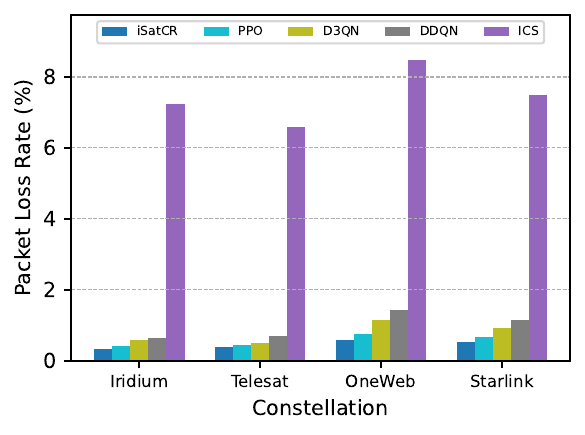}%
        \label{fig:loss}
        }
    \hfill
        \subfloat[]{
        \includegraphics[width=0.315\textwidth]{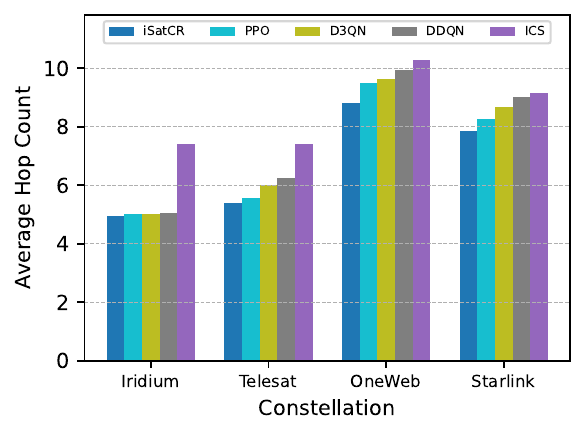}%
        \label{fig:hops}
        }
    \caption{Simulation results of different algorithms under different-scale satellite constellations. (a) Comparison of average task delay. (b) Comparison of packet loss rates. (c) Comparison of average transmission hop count.}
    \label{fig:num}
\end{figure*}

Finally, we evaluated the generalization ability of the proposed method in a test environment different from the training environment. We first kept the task arrival rate constant, while adjusting the link failure rate to 6\%. Tests were conducted in four different-sized satellite constellations: Iridium, Telesat, OneWeb, and Starlink (Gen1 shell2). Fig.~\ref{fig:num} shows the performance of each algorithm in different-sized constellations. It is notable that the proposed iSatCR algorithm consistently maintained the lowest average task delay, average packet loss rate, and average transmission hops across all constellations. The results demonstrate that the iSatCR algorithm has strong robustness and generalization capabilities. This algorithm continuously makes optimal combined computing and routing decisions in complex and variable LEO satellite networks by collecting and analyzing the network topology and resource load data of neighboring satellites for three hops.
 
\section{Conclusions}
\label{sec:conclusion}
This paper investigates a distributed strategy towards integrated computing and routing in the LEO satellite constellation based on graph embedding and DRL. A distributed resource-aware mechanism is proposed based on the graph embedding method with the proposed shifted feature aggregation method. In addition, a D3QN-based method is proposed to achieve a distributed strategy optimizing computing, and routing under the storage limitation. The results of the experiments show the applicability of the proposed iSatCR method, where the overall delay and packet loss can be reduced compared to the baseline methods in high-load scenarios. Furthermore, through simulation experiments conducted under different link failure rates and various constellation scales, it was confirmed that the proposed method possesses strong robustness and generalization ability.

Efforts to improve the performance of the proposed distributed strategy for integrated computing and routing will be pursued in future research. Additionally, efficient models for onboard inference will also be developed.

\bibliographystyle{IEEEtran}
\bibliography{IEEEabrv,reference}

\end{document}